
\input epsf	

\magnification= \magstep1
\tolerance=1600
\parskip=4pt
\baselineskip= 6 true mm

\vsize=7.7in
\hsize=13.3cm
\font\smallrm=cmr9
\font\smallit=cmti9

\font\largerm=cmr12

\def\d{\delta} 		
		 	
\def\m{\mu}	\def\f{\phi} 			\def\vv{\varphi}
\def\n{\nu}	\def\j{\psi} 	
	\def\s{\sigma} 	
	\def\th{\theta}	 	
 			
 		\def\W{\Omega}	

\def\HH{{\cal H}}   \def\NN{{\cal N}}
\def\DD{{\cal D}} \def\AA{{\cal A}} \def\BB{{\cal B}}

\def\cl{\centerline} 	\def\ni{\noindent}
\def\pa{\partial}	\def\dd{{\rm d}}	\def\tt{\tilde}
\def\bra{\langle}	\def\ket{\rangle}

\def\fnd#1{\footnote{$^\dagger$} {\scrunch #1 \toe}}
\def\fndd#1{\footnote{$^\ddagger$} {\scrunch #1 \toe}}
	\def\scrunch{\baselineskip=12 pt \smallrm}
	\def\toe{\hfil\break\vskip-16pt}
	\newcount\noteno
\def\numfn#1{\global\advance\noteno by 1
	\footnote{$^{\the\noteno}$} {\scrunch #1 \toe}}

\def\part#1#2{{\partial#1\over\partial#2}}
\def\on{\over\displaystyle} \def\ref#1{${\vphantom{)}}^{#1}$}
\def\ex#1{e^{\,\textstyle#1}}

\def\qu{\ {\buildrel {\displaystyle ?} \over =}\ }

\def\low#1{{\vphantom{]}}_{#1}}
\def\in{{\,\rm in}}	\def\out{{\,\rm out}}
\def\ret{\hfil\break}

{\vphantom{|}}   \rightline{THU-95/22}\rightline{gr-qc/9509050}

\vfil
{\largerm\ni Quantum information and information loss in General
Relativity
\fndd{This is an elaborated version of a lecture held at the 5th Int.
Symposium on Foundations of Quantum Mechanics in the Light of New
Technology -- {\smallit Quantum Coherence and Decoherence} -- August
21-24, 1995, Adv. Res. Lab, Hitachi, Ltd., Hatoyama, Saitama (Tokyo),
Japan.} }

\vfil
\ni {G. 't Hooft }
\vfil{\ni
 {Institute for Theoretical Physics},
 {University of Utrecht, P.O.Box 80 006} \ret
 {3508 TA Utrecht, the Netherlands}}\ret
 {E-mail: g.thooft@fys.ruu.nl}
\vfil
\ni {ABSTRACT} {\medskip{\scrunch{
When it comes to performing thought experiments with
black holes, Einstein-Bohr like discussions have to be re-opened. For
instance one can ask what happens to the quantum state of a black hole
when the wave function of a single ingoing particle is replaced by an
other one that is orthogonal to the first, while keeping the total
energy and momentum unaffected. Observers at $t\rightarrow\infty$ will
not notice any difference, or so it seems in certain calculational
schemes.

If one argues that this cannot be correct for the complete theory
because a black hole should behave in accordance with conventional
quantum mechanics, implying a unitary evolution, one is forced to
believe that local quantum field theory near the black hole horizon is
very different from what had hitherto been accepted. This would give us
very valuable information concerning physics in the Planck length
region, notably a mathematical structure very close to that of super
string theory, but it does lead to conceptual difficulties.

An approach that is somewhat related to this is to suspect a breakdown
of General Relativity for quantum mechanical systems. It is to some
extent unavoidable that Hilbert space is not invariant under general
coordinate transformations because such transformations add and remove
some states. Finally the cosmological constant problem also suggests
that flat space-time has some special significance in a quantum theory.
We suggest that a new causality principle could lead to further clues
on how to handle this problem.
}\smallskip} }
\vfil\vfil\eject

\ni{\bf 1. INTRODUCTION }\medskip

General relativity and Quantum mechanics are two disciplines in
theoretical physics that have been abundantly tested both
experimentally and from logical mathematical viewpoints. Yet it seems
to be all but impossible to combine the two theories into one. For a
short time it was believed that (super) string theory holds the promise
to answer this need, but one cannot avoid the impression that the
conceptually most difficult issues are not properly assessed by string
theory and that it rather adds its own peculiar interpretation
difficulties to what we already have. As we will briefly explain at a
later stage of this lecture (beginning of Sect. 4), string theory is more
likely to represent some sort of long distance limit, or continuum
limit, of a more detailed theory, as yet ununderstood. In such a
detailed theory space and time may well be discrete in nature, and the
information contained in it must play a central role.

Just because of this discreteness of space and time it is likely that
it will not be possible to make any kind of approximation, or
simplification, so that solvable models would be obtained. Consequently
it may well be that the only way to formulate a correct theory is by
guessing all at once the correct procedure. To make such a guess is
beyond our present capacities. The only alternative as I see it is to
try to argue, as accurately as we can, what the dynamical equations are
likely to be close to the Planck length scale and to try to deduce from
that a picture that is as clear as possible.

This picture should be free from contradictions under all conceivable
circumstances, and this is why it is of importance to consider in
particular the most extreme situations one can imagine. We will
therefore concentrate on the highest possible energy concentrations and
the strongest possible gravitational fields. Quite generally one
expects that the answers one might come forward with will have a
bearing as well on all other related issues in quantum gravity. Indeed,
one answer that seems to emerge from our considerations requires a
reconsideraton of the meaning of quantum theory as a theory describing
reality. More than in previous enterprises where we struggled to find
theories that are appropriately in harmony with quantum mechanics, for
instance the construction of the Standard Model for elementary
particles, the discussion about the interpretation of quantum mechanics
may become relevant\ref1. In fact, the same can be said about General
Relativity\ref2. The extent to which the axioms of this theory will
survive in any ultimate theoretical structure will also have to be
discussed, as we will see in Sects. 5 and 6.  \bigbreak

\ni{\bf 2. BLACK HOLES, HAWKING RADIATION, AND COUNTING STATES}\medskip

As for extreme configurations, our obvious choice is the black hole.
The black hole is a  logically inevitable outcome of the (unquantized)
theory of general relativity. If we  take a sufficiently large quantity
of matter and give it a spherically symmetric initial configuration,
the collapse into a black hole can easily be seen to be unavoidable:
imagine that the total mass would be that of an entire galaxy. In that
case one can deduce that at the moment that the matter particles cross
the horizon (the point of no return near the black hole) the local
density would still be less than that of water, and the laws of nature
needed to describe what happens at that moment are in no way more
exotic than the laws describing water under terrestrial
circumstances\fnd{In practice minute disturbances from the spherical
symmetry of the initial configuration will cause material during its
transition through the horizon to be heated to relativistic
temperatures, which is why in reality black holes often radiate vast
amounts of energy.}. At first sight therefore one would argue that a
correct treatment of physics at the horizon should not lead to any
controversy, and that the real difficulty should be the question what
happens at the origin $r=0$ of the black hole, where a genuine,
physical singularity develops. Curiously, quite the opposite is true.
The singularity at the origin of coordinate space is of no direct
concern since it is safely hidden from our observation, and the
physical relevance of questions about that region is hard to defend.
For all practical purposes one may ignore the singularity. It is the
horizon that gives rise to problems.

Imagine sending an observer into a black hole shortly after the
collapsing material disappeared\ref3. Looking at the solutions to the
equations of motion one finds that this observer will be surrounded by
vacuum. In the simplest, spherically symmetric, case the space-time he
is in is characterized by the Schwarzschild metric: $$\dd
s^2\,=\,-\Big(1-{2M\over r}\Big)\dd t^2+{\dd r^2\on
1-2M/r}+r^2(\dd\th^2+\sin^2\th\,\dd\vv^2)\,, \eqno(2.1)$$ where $M$
stands for $G\low Nm$ and $G\low N$ is Newton's constant, $m$ is the
total mass of the original configuration. One expects that in terms of
his local coordinate frame nothing drastic happens the moment the
observer passes through the horizon, which is at $r=2M$. This is because in
spite of appearances space-time is not singular at that spot. Since we
can compute precisely what the observer will see, we can also deduce
what the outside observer will observe, simply by performing the
relevant coordinate transformation. The first non-trivial calculation
involving quantum field theory here was done by Hawking\ref4 in 1975.
He found that the outside observer will experience thermal radiation at
a temperature $$kT\,=\,{1\over 8\pi M}\,,\eqno(2.2)$$ in Planck units.

One may now imagine doing thermodynamical experiments with a black
hole, injecting and extracting energy while monitoring the
temperature.  The temperature should always be given by Eq. (2.2). This
way one derives\ref{4,5} the {\it entropy} $S$: $$S\,=\,4\pi
M^2\,,\eqno(2.3)$$ again in Planck units. This result seems to make a
lot of sense physically. It suggests that the black hole can be
compared to a macroscopic object such as a balloon filled with a gas
that may escape through an opening. The emission spectrum is thermal as
long as the balloon is macroscopic, but a microscopic desciption, in
terms of a microcanonical ensemble, should be possible also. In such a
system the entropy is directly related to the total number of quantum
states that can describe the system.

\def\stle#1{{#1}\,} One can also try to compute $S$ directly by
counting physical degrees of freedom near the horizon, and it is here
that the first real difficulties show up. Any attempt to count the
physical degrees of freedom near the horizon tends to yield infinity as
an answer\ref6.  Certainly this is what happens if one applies
perturbative field theory near the horizon, treating all Fourier modes
of a field as mutually non-interacting. This is simple to see. Consider
for instance a scalar field $\f$. Let us perform the coordinate
replacement $$r-2M\quad\Rightarrow\quad\ex\s\,,\eqno(2.4)$$ near the
horizon. Then in terms of the coordinates $\s$ and $t$, we have the
Euler-Lagrange field equation:
$$\ddot\f-\stle{1\over(2M)^2}\f_{\s\s}+\ex{\s}\big(-\stle{1\over(2M)^2}
{\pa_{\,\W}}^2\f+m^2\f\big)\,=\,0\,,\eqno(2.5)$$ where ${\pa_{\,\W}}^2$
stands for the angular Laplacian. Closer to the horizon the term with
$e^\s$ may be ignored altogether. Thus we obtain simple plane wave
solutions, but since the boundary is now at $\s=-\infty$ these plane
waves continue their journey towards the horizon without ever returning
-- of course, since this material will vanish into the hole. This
implies that there is a continuum of Fourier modes there, and the heat
capacity of the vacuum here should be strictly infinite, which is not
at all in agreement with the finite expression (2.3) for the entropy.

This infinity problem for the entropy is directly related to the
so-called quantum information problem: one may compare two one-particle
states sent into the black hole, one described by a wave packet that is
orthogonal to the other. These wave packets will continue their
journeys into the hole forever, following the line $\s=-t/2M$, yet it
it difficult to see how they can then affect the outgoing waves, which
will continue to form a thermal spectrum of Hawking particles. The two
in-states could be chosen to be orthogonal to each other, whereas the
out-states are indistinguishable. This appears to violate
unitarity.

The entropy infinity problem is closely related to the (one-loop)
non-renorma\-liz\-ab\-ility of quantum gravity\ref7; the low-energy
contributions to the entropy can be accommodated for by renormalization
of Newton's constant, but this observation does not solve the problem
as it requires an infinite bare Newton constant at a finite (Planckian)
distance scale. See also Ref\ref8.\bigbreak

\ni{\bf 3. INTERACTIONS }\medskip

It is tempting now to suggest that the reason for this apparent
conflict is that we ignored interactions. Now it is true that if we
approach the horizon closer than one Planck length unit ($10^{-33}$ cm)
the effective gravitational interactions become strong there, and
omitting them seems to be a serious mistake. It is however far from
easy to see how one can avoid the disaster just mentioned even when one
does take interactions into account. The local observer sees only
particles going in, nothing comes out. Why would these interact
significantly, and how could this alter the result that the number of
possible states near the horizon diverges to infinity? The particles a
local observer might see coming out seem to be in no way related to the
ingoing objects, and since one is free to choose the initial data for
the ingoing waves during a near infinite amount of time it is hard to
see how nevertheless only a finite amount of independent states would
be possible. To be sure, if one tries to stick to what could be
considered to be well-established rules for applying known laws of
physics, one still does not obtain a finite density of states, even
when one thinks that the interactions {\it have} been taken into
account!

This is why, initially, many researchers indeed found that the density
of states should be infinite\ref9. In thermodynamical language this might
simply mean that the entropy is not given by Eq. (2.3), but by that
with an {\it infinite} constant added to it. This now would have
physically important consequences. It would make the black hole `phase
space' infinite, and simple arguments from quantum scattering theory
then imply that the majority of these states should be absolutely
stable against decay, they would form stable `black hole remnants'.
Remnants are very unlike any known physical objects. Since their
Hilbert space would be strictly infinitely degenerate they would
violate some of the rules of quantum field theory, such as a
recognizable spin statistics connection, and it would be difficult to
see how to avoid infinities in calculations of for instance
gravitational pair production of remnants. Although we realize that
none of these observations are absolute proofs that the remnant theory
is deficient, it is generally considered to be unelegant. It seems to
be more natural to search for a theory that avoids such unconventional
behaviour.

It is certainly possible to find weaknesses in the `naive' arguments
that would point to  the infinite black hole degeneracy. In terms of
locally regular coordinates\ref{10} the horizon can be described as in
Fig. 1.  In terms of the degrees of freedom in a free field theory the
majority of states we are interested in are elements of a product
Hilbert space, $$\HH\,=\,\{|\in\ket\}\times\{|\out\ket\}\,,\eqno(3.1)$$
where the ingoing states $|\in\ket$ include the entire past history and
the outgoing states $|\out\ket$ include the entire future history. This
means that there are infinite amounts of information squeezed onto the
past horizon as well as the future horizon.

\midinsert\epsffile{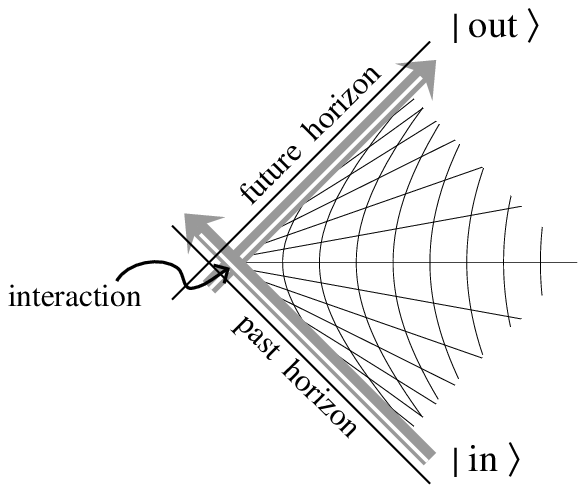}\cl{\scrunch{Fig. 1. The in- and outgoing
particles near the horizon in locally regular coordinates.}}\endinsert

This Hilbert space is larger than may be considered acceptable. In
terms of the regular coordinates as shown in Fig. 1 the ingoing
particles in the far past and the outgoing particles in the distant
future pass each other near the origin with a center-of-mass energy far
beyond the Planck mass, indeed exponentially increasing to beyond the
total energy in the universe. Certainly the mutual gravitational
interactions\ref{11} may then not be ignored. The true Hilbert space is
therefore probably not Eq. (3.1). The crucial question here is how to
deduce what the correct Hilbert space is.

The approach  advocated by us for some time now\ref{12} is to start
from the other end: let us {\it assume} that the black hole Hilbert
space is not infinitely degenerate and that its evolution can be
described by a unitary operator $U(t)$. One can then use conventional
laws of physics to derive constraints that this evolution operator will
have to obey.  This assumption at first sight looks modest and
reasonable but actually turns out to be quite restrictive, with
far-reaching consequences, since this way we can obtain expressions for
the actual form of $U(t)$.  Just because our result is almost in
contradiction with what one would be tempted to derive from standard
laws of physics, some consequences of this assumption imply important
revisions of what one would have thought to be reasonable physics at
the Planck length.  We refer to this approach as {\it the $S$-matrix
Anstaz}.

The $S$-matrix Anstaz only makes sense if we do take interactions
between in- and outgoing matter into account. Most important of all
these interactions is the gravitational one. As has been described in
detail elswhere, the main effect an ingoing particle has on an outgoing
one (and {\it vice versa}) is that its trajectory is shifted\ref{11}.
In terms of the regular local coordinate frame of Fig. 1, let the
ingoing momentum distribution be given by a function $p_\in(\tt x)$,
where $\tt x$ stands for the transverse coordinates, typically the
angles $\th$ and $\vv$ on the horizon. $p_\in$ is actually the
(integrated) energy momentum component $\int\dd x^+T_{++}(\tt
x,x^+,x^-)$ where the continuity equation $\pa_\m T_{\m\n}=0$ removes
nearly all $x^-$ dependence (the spacelike components $T_{+a}$ and
$T_{ab}$ ($a,b\,=\,1$ or $2$) of $T_{\m\n}$ are usually negligible).
Then one finds that the geodesic of an {\it outgoing} particle at
transverse position $\tt x'$ is shifted by an amount $$\d
x^-\,=\,\int\dd^2\tt x f(\tt x-\tt x')p_\in(\tt x)\quad,\qquad f(\tt
z)\,=\,-4G\low N\log|\tt z|\,.\eqno(3.2)$$ The function $f(\tt z)$ can
be seen to be a Green function on the transverse plane, obeying
$${\tilde\pa}^2 f(\tt z)\,=\,-8\pi G\low N\d^2(\tt z)\,.\eqno(3.3)$$
The direction of the shift is in the positive time direction at close
distances, as indicated in Fig. 2.

\midinsert\epsffile{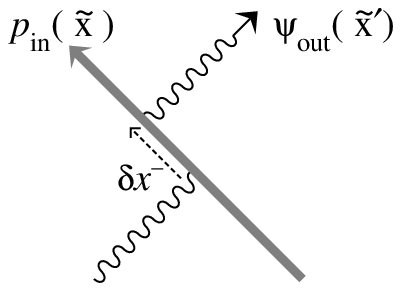}\cl{\scrunch Fig. 2. Shift of a geodesic
due to a fast moving particle.}\endinsert

The effect this has on the wave functions $\j_\out(x)$ is now easy to
compute. In Ref\ref{12} it is shown that the shift (3.2) indeed can be
made compatible with a unitary evolution operator provided that both
the ingoing and the outgoing states are {\it exclusively} characterized
by their momentum distributions $p_\in(\tt x)$ and $p_\out(\tt x)$. In
terms of the Hilbert space elements $|\{p(\tt x)\}\ket$ one finds the
scattering matrix $$\bra\{p_\out(\tt x')\}|\{p_\in(\tt
x)\}\ket\,=\,\NN\ex{-i\int\dd^2\tt x\int\dd^2\tt x'\,p_\in(\tt x)f(\tt
x-\tt x')p_\out(\tt x')}\,,\eqno(3.4)$$ which can be rewritten as the
functional integral $$\bra\{p_\out(\tt x')\}|\{p_\in(\tt
x)\}\ket\,=\,\NN'\int\DD u^\m(\tt x) \ex{\,i\int\dd^2\tt x\big(p_\m(\tt
x)u^\m(\tt x)+{1\over 8\pi G_N}u^+(\tt x){\tilde\pa}^2 u^-(\tt
x)\big)}\,,\eqno(3.5)$$ with $$p_\m(\tt
x)\,=\,(p_+,p_-,0,0)\,=\,(p_\out,-p_\in,0,0)\qquad{\rm and}\qquad
u^\m\,=\,(u^+,u^-,0,0)\,.\eqno(3.6)$$ The quantities $\NN$ and $\NN'$
are normalization factors, to be adjusted such that the matrix (3.4),
(3.5) be unitary.

The result (3.5) can be seen to be remarkably closely related to string
theory amplitudes in spite of the fact that none of the usual premisses
of string theory has been appealed to. The interpretation of (3.5) in
terms of strings has been further explained in Ref\ref{12}. \bigbreak

\ni{\bf 4. COMMUTATION RULES}\medskip

The expressions (3.4), (3.5) are still matrices in an
infinite-dimensional Hilbert space, defined in terms of the continuous
functions $p(\tt x)$. In reality there are reasons to expect only
discrete degrees of freedom. It is important in this respect to
emphasize that in our approach the expressions obtained so-far were
evidently not more than approximations. We suspect that the real
physical degrees of freedom near a black hole horizon must be discrete,
corresponding to one bit of information per Planckian unit of surface
area. Thus, string theory is probably the continuum limit of a discrete
theory.

One should keep in mind that
these expressions were derived assuming that the transverse momenta
were small, otherwise a {\it sideways} gravitational displacement would
have to be taken into account. This means that the expressions are only
to be trusted on a transverse length scale large compared to the Planck
length. One may hope that the introduction of more kinds of
interactions may lead to corrections of this shortcoming, and also to
the expected dimensionality as suggested by the finite entropy (2.3).
Attempts in this direction were made with only partial
success\ref{13,\,14}

In this lecture I now wish to focus on the implications of expressions
of this sort. What has been achieved is that `quantum information',
originally thought to be lost onto the black hole, is actually returned
to us, transformed under the scattering matrix given in some
approximation by (3.4), (3.5). However, there seems to be a price that
has been paid. It is often argued that the result, a non-trivial
unitary evolution law, cannot be correct. These objections were based
on the commutation rules. We will now show why commutation rules in
our scheme have to be handled with great care\ref{15,\,16}.

\midinsert\epsffile{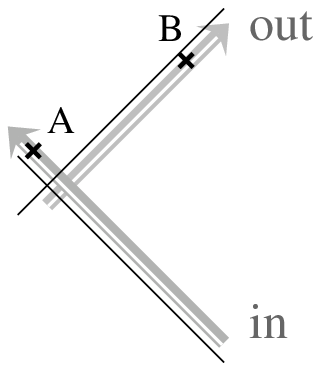}\cl{\scrunch Fig. 3. Considering
operators defined on spacelike seperated points $\scriptstyle A$ and
$\scriptstyle B$.}\endinsert

In Fig. 3. two points near the black hole horizon are indicated, $A$ and
$B$, where operators can be considered: $\AA(x_A)$ and $\BB(x_B)$.
Now one may observe that these points are spacelike sparated, and
therefore one expects
$$[\AA(x_A)\,,\,\BB(x_B)]\,\qu\,0\,.\eqno(4.1)$$
On the other hand however, operator $\AA$ covers all of the Hilbert
space of the ingoing particles, and operator $\BB$ can act on all states
of the Hilbert space of outgoing particles. From this one would deduce
that they cannot always commute. Here we have an apparent paradox.

This is precisely where the $S$-matrix Ansatz deviates from what would
be done in more conventional approaches. The Ansatz implies that the
in- and out-Hilbert spaces are not independent but instead connected by
equations of motion. In terms of the space of states that describe the
Black hole completely and unambiguously the operators $\AA$ and $\BB$
do not commute, but in terms of the states in the corresponding flat
space they do. The situation can be clarified further by regarding
these operators in a Heisenberg picture (i.e. the scheme where
operators are time dependent but states $|\j\ket$ are not) Suppose that
the operator $\AA$ acts on particles that went into the hole in the far
past. This means that these particles in the coordinate frame of Fig. 1
are Lorentz boosted towards the past horizon with boost parameter that
diverges exponentially with time. Particles with this much energy cause
a large gravitational shift among the particles going in the other
direction, that is, the outgoing particles. Thus the operator $\AA$
causes a large deformation of the trajectories of the outgoing
objects.

Now consider the operator $\BB$. One could still maintain that its
effect is independent of what $\AA$ did because it is spacelike
separated. It should be clear however that operator $\BB$ has a similar
effect on the trajectories of the ingoing objects. It is true that in
order to see this effect we have to extrapolate back to the past, but
this is totally legal in the Heisenberg picture. If we were to apply
the conventional argument that $\AA$ and $\BB$ commute this would mean
that we would be working in a Hilbert space where one can create
independently from each other anomalously energetic leftmoving
particles and rightmoving particles. In a Heisenberg picture these
particles would collide with excessive center-of-mass energies near the
origin, causing each other's trajectories to be wildly distorted by
gravitational fields. Since the center-of-mass energy would quickly
exceed the energy needed to form a black hole, we see that black holes,
or rather, their time reverses, sometimes called `white holes', would
become dominant elements of this combined Hilbert space. It is the
philosophy of the $S$-matrix Ansatz that these states would be
unphysical and therefore should be removed from the allowed states. As
soon as one says this however, delicate questions will be raised about
which states should be admitted in Hilbert space, and why. Should one
admit white holes in the far past or not?

It is not unreasonable to speculate that constraints will have to be
introduced to limit oneself to those elements of Hilbert space that
describe in a more or less reasonable way histories of the universe as
we are familiar with. This excludes the states dominated by white
holes. If this is what should be done we can easily sea that $\AA$ and
$\BB$ will not commute. Since we are looking at operators corresponding
to ultra-energetic particles, far outside the realm of conventional
quantum field theory, and since we now do concentrate on {\it the
Hilbert space describing the black hole, and not the Hilbert space of
empty space-time} we think this is legitimate. The argument goes as
follows.

Let us first consider the product $\BB\AA$. The operator $\AA$ is
computed in the point $x_A$ which is after the `observer' crossed the
horizon. One generally argues that the observer did not experience any
special effect while crossing this horizon. In particular his geodesic
was not seriously affected by the gravitational fields of any particle
that might be near him on its way out of the black hole.  After this
one considers that action of oprator $\BB$ which is at a point $x_B$ on
the future horizon. This future horizon is defined by examining the
fate of the outgoing geodesics in the distant future. Therefore the
geodesics of the particles $B$ created by the operator $\BB$ will be
distorted in the region of space-time {\it before} the trajectory of
particle $A$ that went went in.

Now when we compute $\AA\BB$ we may argue that particle $B$ is at the
same spot as before, but what is the point $x_A '$ where one should
elaborate the operator $\AA(x_A ')$? This operator must now be defined
in a spacetime where all trajectories of {\it ingoing} particles $A$
are severely distorted by a strongly gravitating beam of new $B$
particles, that have been added to our state by operator $\BB$. This
now is in direct conflict with the `naive' assertion that the ingoing
observer feels no Hawking particles. The particle $B$ created by the
operator $\BB$ {\it will} be felt by him! Before continuing we should
realize now that we have not yet defined what we exactly mean when we
discuss the operator $\AA(x_A)$: is $x_A$ defined to be the one that is
shifted by the gravitational effects of $B$, or should we take the
point one would have obtained if the observer had indeed not noticed
any shift? It would be reasonable to define $\AA(x_A)$ to be the
operator acting on the point $x_A$ as it would have been if the ingoing
observer had not experienced any gravitational effect from outgoing
particles, since conventional arguments say that such effects will not
be experienced. But id this is the definition we see that in the
product $\AA\BB$ the real point $x_A$ where $\AA$ is calculated is the
shifted one. The most important effect of this shift is that other
particles that crossed the horizon at slightly different transverse
coordinates $\tt x=(\th,\vv)$ will have landed at quite different
locations, so that if $\AA$ depends on {\it different} nearby points
$\tt x$ its effect on these states will be vastly different from what
it did in the product $\BB\AA$. With this definition, $\AA$ and $\BB$
do not commute.  Only if we use the other definition, where the point
$x_A$ is defined in a coordinate frame that had the curvatures due to
the shift caused by $B$ incorporated would $\AA$ and $\BB$ still
commute. This however is a Hilbert space where from the start we
allowed infinitely energetic particles leave the black hole, which in
the far past of its history would be difficult to accept. Such
particles really do not occur in the Hilbert space of the far
past.\bigbreak

\ni{\bf 5. IS GENERAL RELATIVITY VIOLATED?}\medskip

What was learned in the previous chapter is that one should limit
oneself to `reasonable' states in Hilbert space. But what does
`reasonable' mean? If we exclude elementary particles with energies
beyond some large multiple of the Planck energy this may certainly be
called `reasonable', yet of course it violates Lorentz invariance. If
we do not exclude such states we run the danger of allowing states that
have white holes in their past, and this one would tend to call
`unreasonable'. If one is ready to accept the idea of a unitary
evolution of black hole states one must also accept that in the Rindler
frame of Fig. 1, Hilbert space is omly allowed to contain {\it either}
all particle configurations arbitrarily Lorentz-boosted onto the past
horizon (the Hilbert space of the in-states), {\it or} all particle
configurations on the future horizon (the Hilbert space of all
out-states), but not both sets simultaneously, since we expect an
$S$-matrix to connect these two Hilbert spaces. For an observer in a
local inertial frame this situation is unfamiliar. For him leftgoing
particles and rightgoing particles should be allowed to run around
independently, but on the other hand for him particles with energies
much beyond the Planck mass (`trans-Planckian particles') are of little
or no interest. We presently assume that all these Hilbert spaces are
connected via quite non-trivial unitary matrices.

Let us consider again an `observer' falling into the black hole. After
he passed the horizon he might want to make some measurements at the
point $A$ of Fig. 3.  Alternatively we might wish to study the effects
these measurements have on states at the point $B$. The existence of a
unitary evolution matrix should imply that the same measurements could
indeed also have been carried out at $B$. This is reasonable from the
point of view of the black hole physicist, but very strange as
experienced by the observer in the inertial frame. Since $A$ and $B$
seem to be spacelike separated it looks as if we are dealing with the
famous `quantum copying machine' at the interaction point: the state at
$A$ is `duplicated' at $B$.

Even though either the original or the duplicate is guaranteed to
consist of trans-Planckian particles, so that one could maintain that
no conflict has as yet arisen with conventional physics, there is
nevertheless reason to worry. It does not seem unreasonable after all
that trans-Planckian particles are physically realizable. Why then
should these have such odd properties? An answer to this may at first
sight sound like a very radical one, since it will appear to imply that
general relativity is violated. But hang on, I will explain my viewpoint
about this afterwards.

The state at $A$ is not duplicated at $B$, but rather transformed into
it:  there is no way of describing a Hilbert space that contains both
all possible states in $A$ and all possible states in $B$. Thus, we
must {\it choose} which of these two representations of Hilbert space
we wish to use.  A simple model explains the situation best. Consider a
quantum system described by a Hamiltonian, $$ \eqalign{{\rm at}\quad
t\le 0\,:\qquad& H(t)\,=\,H_0\,,\cr {\rm at}\quad
t>0\,:\qquad&H(t)\,=\Bigg\{ \matrix{H_1(t)\ {\rm
in\ universe\ \#1}\,,\vphantom{\Big)}\cr H_2(t)\ {\rm
in\ universe\ \#2}\,.\vphantom{\Big)}\cr}\cr}\eqno(5.1)$$ Thus at $t>0$
we have two Hamiltonians to choose from. A unitary transformation
relates the two branches, but the existence of such a transformation is
a mere formality. For all we know the two `worlds' at $t>0$ are
different. In a Heisenberg picture, operators in universe \#1 at $t=+1$
sec do not commute with operators in universe \#2 at $t=+1$ sec. Yet
information cannot be readily transmitted from 1 to 2. Our two
Hamiltonians could refer to the two cases: an observer does or does not
pass through the horizon.

The danger of theories of this sort is also clearly exhibited in this
model. There needs not be any relation between $H_1(t)$ and $H_2(t)$.
Indeed the same situation would occur if the theory of General
Relativity that normally relates the two universes were completely
violated. If our model would refer to a black hole one could interpret
is as follows: {\it one} Hamiltonian describes the black hole for {\it
all} observers, the other would describe {\it flat space only}.
``Reality" is only described by {\it one} Hamiltonian. This is why our
model is equivalent to a theory with explicit violation of General
Relativity.

We therefore have to deal with the problem how to explain the accurate
validity of General Relativity for weak gravitational fields. It must
be possible to deduce the efficiency of General Relativity for the
ordinary world, from some symmetry principle. This symmetry principle
should also mean something for Rindler space. We believe that even
though $H_1$ and $H_2$ need not be the same, there should still exist a
well-defined transformation between them. This transformation law could
however be much more complicated than the ones usually employed in
General Relativity. In short: a symmetry principle that implies General
Relativity in the classical limit may still exist, but Hilbert space
itself cannot be completely invariant under coordinate
reparametrizations. Our big challenge is now to identify the correct
symmetries and transformation rules.\bigbreak

\ni{\bf 6. CAUSALITY}\fnd{This chapter regrettably could not be
discussed at the Conference for lack of time.}\medskip

If one would be ready to relax the demands of General Relativity in our
description of the black hole horizon, other fundamental principles
might have to be called for. A very fundamental principle in physics,
often rudely disregarded, is causality. It is true that this demand has
become more questionable, and it has become a fashion to reject it
altogether. As the space-time metric  $g_{\m\n}(x)$ has become a
quantum variable itself it seems to be impossible to keep the speed of
light as an upper bound of the velocity of information transport. Often
it is suspected that there are contributions in the functional
integrand where $g_{\m\n}$ has lost its Lorentzian signature\ref{17}.
Worse even, topologically non-trivial excitations such as wormholes may
produce space-times harboring closed timelike loops\ref{18}. Such
proposals completely ignore the demand that there should be a strict
separation between cause and effect, not only in a classical system but
also in the Schr\"odinger equation.

My proposal is to restore causality, and it may even prove to be useful
to restore the speed of light as an upper bound. First let me explain
that in classical general relativity the {\it coordinate speed of
light} is indeed an upper bound for information transport. More
precisely, one has a theorem of the following
sort:{\smallskip{\narrower\ni\it In any gravitating system surrounded
by asymptotically flat space, with a $T_{\m\n}$ satisfying a positive
mass-energy condition, a coordinate frame $\{x^\m\}$ can be found such
that:\hfil\break 1) $x^\m$ approaches the asymptotically flat
coordinates $x_0^\m$ sufficiently rapidly, and\hfil\break 2) for all
infinitesimal $\dd x^\m$ one has $$g_{\m\n}\dd x^\m\dd
x^\n\,\ge\,\eta_{\m\n}\dd x^\m\dd x^\n\,,\eqno(6.1)$$ with
$$\eta_{\m\n}\,=\,{\rm diag}\,(-1,1,1,1)\,.\eqno(6.2)$$}\smallskip}\ni
This ensures that the dynamical lightcone resides inside the coordinate
lightcone. The proof of the general theorem is as yet in a stage of a
conjecture, but it is easy to verify for the Schwarzschild,
Reissner-Nordstrom, Kerr and Kerr-Newman solutions. The theorem is
illustrated by the well-known Shapiro delay of radar signals passing
close by the Sun. It would not hold for negative Schwarzschild masses.
The boundary condition on the coordinate frame has to be formulated
with sufficient care. One may for instance demand that $x^\m$ rapidly
approaches the usual Schwarzschild coordinate frame.

One may postulate this theorem now also to hold for the quantum case.
This means that one is forced to use a particular coordinate frame.
Transition towards any other frame will not be forbidden, but ensues in
the admission of new states in Hilbert space (ones allowing values of
the quantum variable $g_{\m\n}(x)$ that had been forbidden in the
previous frame), while it will force the removal of others. This will
be in accordance to the theme of the previous section where we describe
how the Rindler space transformation adds and removes some elements in
Hilbert space. Recognising that flat space-time at the boundary plays
an essential role in formulating the quantum theory may well lead to an
additional bonus: the resolution of the well-known cosmological
constant problem may be impossible without the use of flat space-time
as a special reference system where the energy density must be tuned to
zero.

My causality postulate may perhaps not be welcomed by investigators who
speculate on the occurrence of phenomena such as topology change, but
the apparent validity of our theorem for classical gravity may provide
some new thoughts.\bigbreak

\ni{\bf REFERENCES}\medskip

\item{1} G. 't Hooft, {\it J. Geometry and Physics} {\bf 1} (1984)
45-52; see also G. 't Hooft, {\it Black Holes and the Foundations of
Quantum Mechanics}, in ``Niels Bohr: Physics and the World", ed. H.
Feshbach et al (Harwood Acad. Publ., London, Paris, New York,
Melbourne, 1988), p. 171.

\item{2} G. 't Hooft, {\it J. Stat. Phys.} {\bf 53}, (1988) 323.

\item{3} S.W.  Hawking and  G.F.R.  Ellis, {\it The Large Scale
Structure of Space-time}, Cambridge Univ. Press (Cambridge, 1973).

\item{4}   R.M. Wald, {\it Commun. Math. Phys.} {\bf 45} (1975) 9; S.W.
Hawking, {\it Comm. Math. Phys. }{\bf 43} (1975) 199; J.B. Hartle and
S.W. Hawking, {\it Phys.Rev.} {\bf D13} (1976) 2188; W.G. Unruh, {\it
Phys. Rev.} {\bf  D14} (1976) 870.

\item{5} R.M. Wald, {\it Phys. Rev.} {\bf  D20} (1979) 1271; J.D.
Bekenstein, {\it Nuovo Cim. Lett.} {\bf 4} (1972) 737, {\it Phys. Rev.}
{\bf D5} (1972) 1239, 2403; {\bf  D7} (1973) 2333; {\bf  D9} (1974)
3292; {\it Nuovo Cim. Lett.} {\bf 4} (1972) 737.

\item{6} G. 't Hooft, {\it Nucl. Phys.} {\bf  B256} (1985) 727.

\item{7} J.L.F. Barb\'on and R. Emparan, {\it On quantum black hole
entropy and Newton constant renormalization} , Princeton and Bilbao
prepr. PUPT-1529 / EHU-FT 95/5, hep-th/9502155.

\item{8} S. Carlip, {\it Statistical Mechanics and Black Hole Entropy}
UCD-95-30, gr-qc/9509024, September 1995.

\item{9}  C. Callan, S. Giddings, J. Harvey and A. Strominger,{\it
Phys.  Rev.} {\bf D45} (1992) 1005.

\item{10} W. Rindler, {\it Am.J. Phys.} {\bf 34} (1966) 1174.

\item{11} P.C. Aichelburg and R.U. Sexl, {\it Gen.Rel. and Gravitation}
{\bf 2} (1971) 303; T. Dray) and G. 't Hooft,  {\it Nucl. Phys. }{\bf
B253} (1985) 173; {\it Comm. Math. Phys.} {\bf 99} (1985) 613.

\item{12}  G. 't Hooft, {\it Phys. Scripta} {\bf T15} (1987) 143; {\it
Nucl.  Phys.} {\bf B335} (1990) 138; {\it Acta Physica Polonica} {\bf
B19} (1988) 187; {\it Strings from gravity} , In:{\it  Unification of
Fundamental Interactions}.  Proceedings of Nobel Symposium 67,
Marstrand, Sweden, June 2-7, 1986. Eds.  L. Brink et al.; {\it Physica
Scripta,} {\bf T15} (1987) 143; {\it Scattering matrix for a quantized
black hole} , In book: {\it Black Hole Physics}, ed.  V. De Sabbata and
Z. Zhang . 1992 (Kluwer Academic Publishers, The Netherlands), p. 381.

\item{13} G. 't Hooft, {\it Nucl. Phys. }{\bf B43} (Proc. Suppl.)
(1995) 1; {\it Physica Scripta} {\bf T36} (1991) 247.

\item{14} G. 't Hooft, {\it More on the black hole $S$-matrix} , In:
Proceedings of the Fifth Seminar {\it Quantum Gravity}, Moscow, USSR,
28 May-1 June 1990, ed. M.A. Markov, V.A. Berezin and V.P. Frolov.
World Scientific, Singapore (1991) p.  251; {\it S-Matrix theory for
black holes} , in Proceedings of a NATO Advanced Study Institute on New
Symmetry Principles in Quantum Field Theory, held July 16-27, 1991 in
Carg\`ese, France. Eds. J. Fr\"ohlich, G. `t Hooft, A. Jaffe, G. Mack,
P.K. Mitter and R. Stora NATO ASI Series  (1992 Plenum Press, New York)
p.  275.

\item{15} C.R. Stephens, G. 't Hooft and B.F. Whiting, {\it Class.
Quantum Grav.} {\bf 11} (1994) 621; E. Verlinde, {\it Black Hole
Evaporation and Complementarity} , to appear in the proceedings of the
1993 Trieste Spring School,  hep-th/9503120.

\item{16} L. Susskind, L. Thorlacius and J. Uglum, {\it Phys. Rev.}
{\bf D48} (1993) 3743 (hep-th 9306069); D.A. Lowe {\it et al}, {\it
Black Hole Complementarity {\it vs.} Locality} , prepr. NSF-ITP-95-47 /
SU-ITP-95-13 ; {\it Phys. Rev.}   {\bf D48} (1993) 3743, ibid. {\bf
D49} (1994) 966;  D.~A.~Lowe, L.~Susskind, and J.~Uglum, {\it Phys.
Rev.}  {\bf B327} (1994) 226; L.~Susskind, {\it The World as a
Hologram}, preprint SU-ITP-94-33, hep-th/9409089.

\item{17} J.B. Hartle, {\it J. Math. Phys.} {\bf 30} (1989) 452.

\item{18} J.R. Gott, {\it Phys. Rev. Lett.} {\bf 66} (1991) 1126; S.W.
Hawking, {\it Phys. Rev.}  {\bf D46} (1992) 603; C.J. Fewster and C.G.
Wells, {\it Unitarity of Quantum Theory and Closed Timelike Curves} ,
Cambridge prepr. DAMTP-R94/35, hep-th/9409156.

\end